# Existence and interpretation of smooth cosmic-ray dominated shock structures in supernova remnants

L O'C Drury[1], H J Völk[2] and E G Berezhko[3]

[1] Dublin Institute for Advanced Studies, 5 Merrion Square, Dublin 2, Ireland.
[2] Max-Planck-Institut für Kernphysik, Postfach 103980, Heidelberg, Germany.
[3] Institute of Cosmophysical Research and Aeronomy, Lenin Ave 31, 677891 Yakutsk, Russia.

**Abstract.** Studies of energetic charged particle acceleration at collisionless shocks by the diffusive shock acceleration process indicate that completely smooth cosmic ray dominated shock structures are possible. The significance of these is discussed and conditions for their existence derived. It is shown that such solutions cannot dominate the evolution of supernova remnants if the particle diffusion coefficient has the expected strong momentum dependence.

## 1. Introduction

The early analyses of steady planar shock structures incorporating energetic particle (cosmic ray) acceleration demonstrated that, in addition to the expected solutions with a modified gas subshock and an extended cosmic ray precursor, there also existed solutions which were completely smooth (Drury and Völk 1981, Axford et al 1982, Achterberg et al 1984). In the terminology introduced by Draine for molecular shocks these are C-type cosmic ray shocks (C for continuous, as distinct from J-type shocks which contain a "jump" in the hydrodynamic quantities). Because the entire irreversible dissipation of mechanical energy in these shocks is mediated by and goes into the cosmic ray population they have also been called the "efficient" solutions.

Such solutions are completely reasonable and understandable in cases where there is a significant pre-existing population of cosmic rays in the medium into which the shock is propagating. The worrying feature is that such solutions also exist even when there are no up-stream cosmic rays. Here a cosmic ray population seems to appear out of nothing and take over the shock; certainly a rather unphysical picture. Indeed the physical validity of these solutions has often been questioned (*eg* Jones and Ellison, 1991). We note that similar problems do not arise for the J-type solutions with a gas subshock, even though these can also have cosmic ray precursors with no preexisting upstream cosmic rays, because it is expected that the gas subshock will continuously "inject" newly created suprathermal particles into the acceleration process. Of course at a deeper level of description the J-type subshocks also have a structure which is in principle resolved in theories where no artificial distinction is made between thermal and non-thermal particles. The assumption we are making here is simply that the injection of particles from the thermal population and their subsequent acceleration to supra-thermal energies requires the velocity profile to have gradients of order a few thermal ion gyroradii, which at the level of a hydrodynamic description are then equivalent to jump discontinuities in the velocity.

The question is of more than purely academic interest. If the shocks bounding supernova remnants (SNRs) can become cosmic ray dominated C-type shocks, this has serious implications for the interpretation of SNR shock diagnostics (Boulares and Cox, 1988). In the extreme case, were the shocks to become almost immediately C-type, one would have difficulty understanding the X-ray observations which clearly show that SNRs contain shock-heated gas (eg Aschenbach, 1993). Previous work (Drury et al 1989, Markiewicz et al, 1990) indicated that the tendency of shocks to become C-type was less in expanding spherical shocks than in steady planar models; however at high values of the particle injection these "simplified models" still became cosmic ray dominated. Recently much improved models have been calculated by Berezhko et al (1994) which show that SNR shocks remain J-type even at very high injection rates. In an attempt to interpret these results we have been lead to reconsider the significance of C-type solutions and the conditions under which they can exist.

## 2. Interpretation

The only satisfactory interpretation of the C-type solutions with no cosmic-ray source is, we believe, that outlined in Drury (1983). If there is no current source of particles, they must have been introduced in the past. A fundamental feature of diffusive shock acceleration is that some few particles spend a very long time in the neighbourhood of the shock (and in the process get accelerated to very high energies). Thus it is possible for a period of injection in the distant past to leave behind a residual population of particles which, although ever decreasing in number, are still gaining energy at such a rate that the energy density of the particles can remain constant.

In this note we wish to expand slightly on the analysis of this interpretation given in Drury (1983) and to emphasise some interesting new geometrical aspects which are relevant to the spherical SNR case.

The acceleration time scale for particles of momentum $p$ is given by the well known formula,

$$t_{\rm acc} = \frac{3\,L(p)}{\Delta U}$$

where $L(p)$ is the size of the region around the shock occupied by the particles which are being accelerated and $\Delta U$ is the velocity difference across the shock. Normally one assumes

$$L(p) = \frac{\kappa_1(p)}{U_1} + \frac{\kappa_2(p)}{U_2}$$

where the subscripts 1 and 2 refer to upstream and downstream values of the velocity $U$ and diffusion coefficient $\kappa$.

Let us suppose that the diffusion coefficient has a power-law dependence on particle momentum, $\kappa(p) \propto p^\alpha$. The physically significant case is probably $\alpha \approx 1$, but most analytical work has assumed $\alpha = 0$ and for technical reasons a lot of numerical work has used small values of $\alpha$, for example $\alpha = 0.25$ was used by Falle and Giddings (1987). The first reported calculations with $\alpha = 1$ were those of Duffy (1992). Note that strictly stationary shock structures (eg those considered by Drury and Völk, 1981) automatically imply $\alpha = 0$.

The analysis in Drury (1983) consisted in looking at a planar shock with constant velocity jump $\Delta U$. We then have $t_{\rm acc} \propto L \propto \kappa \propto p^\alpha$ and thus

$$\frac{dp}{dt} = \frac{p}{t_{\rm acc}} \propto p^{1-\alpha}$$

so that $p \propto t^{1/\alpha}$. In the limit $\alpha = 0$ this power-law dependence becomes an exponential and $p \propto \exp t$.

Let us ignore for the moment the reduction in the number of particles by advection downstream away from the shock. Then we have a fixed number of particles all gaining energy so that $p \propto t^{1/\alpha}$. However these are confined to a region of width $L(p)$ about the shock where, as is easily seen, $L \propto t$. Thus in planar geometry the energy density, or equivalently the pressure, of the accelerated particles scales as $t^{(1-\alpha)/\alpha}$ and can increase with time as long as $\alpha < 1$. Thus by allowing the correct amount of downstream advection it is possible to have planar C-shocks with constant pressure and no fresh particle injection. Note that the smaller the value of $\alpha$ the easier it is to have such solutions.

The interesting new feature to which we wish to draw attention in this note is what happens in spherical gometry. If the shock is moving at constant speed (as for example a SNR shock during the sweep-up phase) the above analysis applies except that in addition to the widening of the acceleration zone thickness $L$ there is a geometrical expansion of the shock area by a factor $R^2 \propto t^2$. Thus the volume in which the accelerated particles are dispersed increases as $t^3$ and the pressure scales as $t^{(1-3\alpha)/\alpha}$. We conclude that self-sustaining C-type solutions are impossible unless $\alpha < 1/3$ during the free expansion phase.

In general, for a spherical shock moving at non-constant speed, the condition that a self-sustaining C-shock be possible is that the cosmic ray pressure remain capable of balancing the ram pressure, that is

$$\delta\left(p_{\rm max} R^{-2} L^{-1}\right) > \delta\left(\rho \dot{R}^2\right)$$

where $\delta$ denotes the logarithmic derivative, $\delta X = \dot{X}/X$. This can be written as

$$\delta p_{\rm max} - 2\delta R - \delta L > \delta \rho + 2\delta \dot{R}$$

or noting that $\delta p_{\rm max} = 1/t_{\rm acc}$ (by definition) and that $\delta L = \alpha \delta p - \delta R$,

$$\frac{1-\alpha}{t_{\rm acc}} > \delta \rho + 2\delta R + \delta \dot{R}$$

Assuming a power-law dependence of the radius on time, $R \propto t^\beta$, this can be written

$$\frac{1-\alpha}{t_{\rm acc}} > \frac{3\beta - 1}{t} + \delta \rho.$$

We see that self-sustaining C-type shocks are also impossible for Sedov-like SNR shocks with $t_{\rm acc} \approx t$, $\alpha = 1$, $\beta \approx 2/5$ and $\delta \rho = 0$. However they may be possible for shocks propagating outwards in a region of rapidly decreasing density. Specifically, for a shock propagating in a region where the density falls as $R^{-2}$, as expected in many wind solutions, $\delta \rho = -2\beta/t$, and the condition becomes marginal. However in this case one must also allow for the variation in the magnetic field with radius and its effect on the diffusion coefficient as well as the relative contributions of parallel and perpendicular diffusion.

## 3. Conclusions

The somewhat surprising result of this elementary analysis is that SNR shocks cannot be cosmic-ray dominated over long periods of the SNR evolution, in agreement with the recent calculations of Berezhko et al. The cosmic ray dominated shocks which formed at high injection rates in the simplified model calculations resulted, we believe, from the particular form of effective diffusion coefficient used in these calculations. The effective diffusion coefficient was taken to be of order $\kappa(p_{\rm max})/\ln(p/mc)$ which is correct as long as one has a roughly power-law spectrum stretching from sub-relativistic energies to the maximum momentum $p_{\rm max}$. However if the subshock disappears and the injection stops, all of these low-energy particles are rapidly accelerated up to close to $p_{\rm max}$ and the effective diffusion coefficient should rapidly increase from a value of order $\kappa(p_{\rm max})/10$ to a value close to $\kappa(p_{\rm max})$. Because this effect was overlooked C-type shocks, once established in the calculations, tended to persist until quite late in the Sedov phase. We emphasise that there is nothing inherently wrong with the simplified models, nor the two-fluid approximation on which they are based. In fact we believe such analyses are very helpful in understanding detailed calculations such as those by Berezhko et al. It is however unfortunately true, as has often been emphasised, that they are very sensitive to the values of the closure parameters $\gamma_c$ (adiabatic exponent of the cosmic rays) and $\langle \kappa \rangle$ (effective diffusion coefficient) used.

Finally we note that these results strengthen the theoretical case for shock acceleration in SNRs as the source of the bulk of the Galactic cosmic rays.

*Acknowledgements.* LD and EB gratefully acknowledge the hospitality of the Max-Planck-Institut für Kernphysik where this work was carried out.